\documentclass[12pt,a4paper]{article}
\usepackage{a4wide,epsf}
\usepackage[latin1]{inputenc}
\usepackage{latexsym}
\newcommand{\dddot}[1]{\stackrel{...}{#1}}
\newcommand{\pkt}{\; .}
\newcommand{\kma}{\; ,}

\newcommand{\re}[1]{{\rm Re }\,#1}

\newcommand{\calF}{{\cal F}}
\newcommand{\calC}{{\cal C}}
\newcommand{\calS}{{\cal S}}
\newcommand{\calE}{{\cal E}}

\newcommand{\intk}{\int\frac{d^3k}{(2\pi)^3 2\omega_{k0}}\,}
\newcommand{\bfk}{{\bf k}}
\newcommand{\bfx}{{\bf x}}
\newcommand{\be}{\begin{equation}}
\newcommand{\ee}{\end{equation}}
\newcommand{\bea}{\begin{eqnarray}}
\newcommand{\eea}{\end{eqnarray}}
\newcommand{\omk}{\omega_{k0}}
\begin{document}
\begin{titlepage}
\begin{flushright}
DO-TH-97/24\\
November 1997
\end{flushright}
\vspace{20mm}
\begin{center}
{\Large \bf
On the choice of initial states in nonequilibrium dynamics }

\vspace{10mm}
{\large  J\"urgen Baacke\footnote{
 e-mail:~baacke@physik.uni-dortmund.de}, Katrin Heitmann
\footnote{e-mail:~heitmann@hal1.physik.uni-dortmund.de}, and
Carsten P\"atzold
\footnote{e-mail:~paetzold@hal1.physik.uni-dortmund.de}} \\
\vspace{15mm}

{\large Institut f\"ur Physik, Universit\"at Dortmund} \\
{\large D - 44221 Dortmund, Germany}
\vspace{25mm}

{\bf Abstract}
\end{center} 
Imposing initial conditions to nonequilibrium systems
at some time $t_0$ leads, in renormalized quantum field theory,
to the appearance of singularities in the
variable $t-t_0$ in relevant physical quantities, such as
energy density and pressure.
These ``initial singularities'' can be traced back to the choice of
initial state. We construct here, by a Bogoliubov transformation,
initial states such that these singularities are eliminated. 
While the construction is
not unique it can be considered  a minimal way of taking into 
account the nonequilibrium evolution of the
system prior to $t_0$.
\end{titlepage}


\section{Introduction}
Nonequilibrium dynamics in quantum field theory 
has become, during the last years, a
very active field of research in particle physics [1-6],
in cosmology [7-20]
, and in solid state physics \cite{Zurek:1996}.
The outline of typical computational
experiments is as follows: a quantum field $\psi(\bfx,t)$ is driven
by a classical field degree of freedom (Higgs, inflaton, condensate)
$\phi(t)$ which takes an initial value away from
a local or global minimum of the classical or effective action;
the time development is then studied including the
 back reaction of the quantum field in one-loop, Hartree or
large-N approximations. The
initial state of the quantum field $\psi$ is usually taken to be
the vacuum state corresponding to a free field of some
``initial mass'' $m(t_0)$ or a thermal state built on such 
a vacuum state.
In  Friedmann-Robertson-Walker (FRW) cosmology 
\cite{Boyanovsky:1997a,Ramsey:1997,Baacke:1997c}
the usual initial quantum state 
is chosen to be the  conformal vacuum, again corresponding to the
initial mass $m(t_0)$.
While such choices seem very natural, they are not necessarily
appropriate; nevertheless, this point has received little attention
up to now. The reason why we address this question is the occurrence,
in some dynamically relevant physical quantities,
of singularities in the time variable which are related
to the choice of initial state. We will in fact show that
these singularities can be removed by more appropriate choices.

The origin of these singularities can be traced back to
a discontinuous switching on of the interaction with the
external field $\phi(t)$.
This interaction is given by a time-dependent mass term \footnote
{We choose $t_0=0$ for convenience.} 
$m^2(t)=m^2(0)+V(t)$ where for the simplest case of a 
$\lambda\Phi^4$ theory \cite{Boyanovsky:1994,Baacke:1997a}
\be
V(t)=\frac{\lambda}{2} \left [\phi(t)-\phi(0)\right]
\pkt\ee
In FRW cosmology the (conformal) time dependent mass term reads
\cite{Ringwald:1987,Baacke:1997c}
\be
\label{effm}
M^2(\tau)=a^2(\tau)\left\{ m^2+\left(\xi-\frac 1 6 \right)R(\tau)+\frac
\lambda 2 \frac{\varphi^2(\tau)}{a^2(\tau)}
\right\}\pkt
\ee
In this case $M^2(\tau)$ and therefore 
$V(\tau)= M^2(\tau)-M^2(0)$ also contain the scale parameter
$a(\tau)$ and the curvature scalar $R(\tau)$.

A free field theory vacuum state  
corresponding to the mass $m(0)$ would be an appropriate
equilibrium state if $V(t)$ stayed zero for all times.
However, at $t=0$ the potential $V(t)$ changes in a nonanalytic way.
This is unavoidable since at least the second derivative of
$\phi(t)$ becomes nonzero on account of the equation of motion.
As a result of these discontinuities
relevant physical quantities develop singularities in the time variable
$t$ at $t=0$. In the case of $\lambda\Phi^4$ theory in 
Minkowski space
such singularities only occur in the pressure. 
In FRW cosmology the problem becomes more acute.
There,  even the first derivative of
$V(t)$ necessarily becomes nonzero at $t=0$; indeed, even with
a constant external field $\phi$ the initial state could not be at
equilibrium; this manifests itself by a nonvanishing
 first derivative of the scale parameter induced
by the Friedmann equations. 
Furthermore, in this case
both energy and pressure become singular; since they
 enter the Friedmann equations this singular
behavior also affects the dynamics.

Singularities arising from imposing initial conditions to
quantum systems have been noted
for the first time by St\"uckelberg \cite{Stueckelberg:1951},
who called them ``surface singularities''; they are 
briefly mentioned in the textbook of Bogoliubov and Shirkov
\cite{Bogoliubov:1980}.
The ``Casimir effect'' arising from initial conditions has been
discussed by Symanzik \cite{Symanzik:1981}.
In the context of nonequilibrium
dynamics in FRW cosmology the occurrence of such singularities
has been noted by Ringwald \cite{Ringwald:1987}.
In the following we will refer to these singularities 
as ``initial singularities''.

Imposing an initial condition at some time $t_0$
does not mean, in most
applications, that one assumes the system to have come
into being at just this time. Rather, $t_0$ is 
usually chosen as a point in time at which one
can make, on some physical grounds, plausible assumptions
about the state of the system. Clearly, if the system is
not at equilibrium after $t_0$, it will not have been so
before. Therefore, the initial state should take into account,
at least in some minimal way, the previous nonequilibrium evolution
of the system. Such a minimal requirement is the vanishing
of initial singularities. It is the aim of this paper to
specify such initial states.

The plan of the paper is as follows: in section 2
we present the basic
equations and formulate the problem
for the case of a $\lambda\Phi^4$ theory;
in section 3 we discuss an appropriate choice of the
initial quantum state such that the singular behavior
in the time variable is removed;
the modified renormalized equations for the
nonequilibrium system are given in section 4; 
we end with some concluding remarks in section 5.


\section{Formulation of the problem}
\setcounter{equation}{0}
 We consider a scalar $\lambda\Phi^4$ theory without
spontaneous symmetry breaking. The Lagrangian density is given by
\begin{equation}
{\cal L}=\frac{1}{2}\partial_\mu\Phi\partial^\mu\Phi
-\frac{1}{2}m^2\Phi^2-
\frac{\lambda}{4!}\Phi^4.
\end{equation}
We split the field
$\Phi$ into its expectation value $\phi$ and the quantum fluctuations
$\psi$:
\begin{equation}
\label{erw}
\Phi(\bfx,t)=\phi(t)+\psi(\bfx,t)\kma
\end{equation}
with
\begin{equation}
\phi(t)=\langle\Phi(\bfx,t)\rangle=
\frac{{\rm Tr}{\Phi\rho(t)}}{{\rm Tr}\rho(t)}\kma
\end{equation}
where $\rho(t)$ is the density matrix of the system which satisfies
the Liouville equation
\begin{equation}
i \frac{d\rho (t)}{dt} =  [{\cal H}(t),\rho(t)] \; .
\end{equation}
The Lagrangian then takes the form
\begin{equation}
{\cal L}={\cal L}_{\rm 0}+{\cal L}_{\rm I}\kma
\end{equation}
with
\begin{eqnarray}
{\cal L}_{\rm 0}&=&\frac{1}{2}
\partial_\mu\psi\partial^\mu\psi-\frac{1}{2}m^2\psi^2\nonumber\\
&&+\frac{1}{2}
\partial_\mu\phi\partial^\mu\phi-\frac{1}{2}m^2\phi^2-
\frac{\lambda}{4!}\phi^4 \; ,\\
{\cal L}_{\rm I}&=&
\partial_\mu\psi\partial^\mu\phi-m^2\psi\phi-
\frac{\lambda}{4!}\psi^4-\frac{\lambda}{6}\psi^3\phi
-\frac{\lambda}{4}\psi^2\phi^2-\frac{\lambda}{6}\psi\phi^3 \; .
\end{eqnarray}

The equation of motion for the field $\phi(t)$
is given by \cite{Boyanovsky:1994}
\begin{equation} \label{phidgl}
\ddot{\phi}(t)+m^2\phi(t)+\frac{\lambda}{6}\phi^3(t)
+\frac{1}{i}\frac{\lambda}{2}\phi(t) G^{++}(t,\bfx;t,\bfx)=0\pkt
\end{equation}
Here $G^{++}$ is the $++$ matrix element of 
the exact nonequilibrium 
Green function \cite{Schwinger:1961,Keldysh:1964}
in the background field
$\phi(t)$. 
For a pure initial state $|i\rangle$ it can be written as
\be
-iG^{++}(t,\bfx;t',\bfx')=
\langle i| T\psi(t,\bfx)\psi(t',\bfx')|i \rangle \pkt
\ee
If the classical field is spatially uniform
the equation of motion for the field $\psi(t,\bfx)$
is given by
\be
\left[\frac{\partial^2}{\partial t^2}-
\Delta+m^2+\frac{\lambda}{2}\phi^2(t)\right]
\psi(t,\bfx) = 0 \pkt
\ee
We introduce the notations
\bea \label{massoft}
m^2(t)&=&m^2+\frac{\lambda}{2}\phi^2(t)\kma \\
\label{omoft}
\omega_k(t)&=&\left[ \bfk^2+
m^2(t)\right]^{\frac{1}{2}}\kma
\eea
and
\begin{equation}
\omk=\left[ \bfk^2+m_0^2
\right]^\frac{1}{2} \pkt
\ee
We will discuss the choice of $m_0$ below.
We define the `potential' $V(t)$ as
\be
V(t)=\omega_k^2(t)-\omk^2
\ee
We further introduce the mode functions for fixed momentum
$U_{k}(t)\exp(i\bfk\cdot\bfx)$
\footnote{Note that the functions $U_{k}(t)$ depend only on the
absolute value of $\bfk$.} which satify the evolution
equation
\be
\left[\frac{\partial^2}{\partial t^2}+
 \omega_k^2(t)\right] U_k(t)=0 \; ;
\ee
we choose the initial conditions
\be
U_k(0)=1 \hspace{1cm} \dot U_k(0)=-i\omk \pkt
\ee
The field $\psi$ can now be expanded as
\be \label{fieldexp}
\psi(t,\bfx)=\intk
\left[a(\bfk)U_{k}(t)e^{i\bfk \cdot \bfx}
+a^\dagger(\bfk)U^*_{k}(t)e^{-i\bfk \cdot \bfx}
\right]\kma
\ee
where the operators $a(\bfk)$ satisfy
\be
[a(\bfk),a^\dagger(\bfk')]=(2\pi)^3 2\omk
\delta^3(\bfk -  \bfk')
\ee
If the initial state $|i\rangle$ is chosen as
the vacuum state corresponding to the operators
$a(\bfk)$, i.e., as satisfying $a(\bfk)|i\rangle =0$,
 we obtain the Green function
$G^{++}(t,t';\bfx-\bfx')$ as
\be
G^{++}(t,t';\bfx -\bfx')=\intk
\left[ U_k(t)U^*_k(t')\theta(t-t')
+U_k(t')U^*_k(t)\theta(t'-t)\right] e^{i\bfk (\bfx -\bfx')}
\pkt
\end{equation}
The Green function at equal space and time points then reads 
\be
G_k^{++}(t,t;{\bf 0})=
i \intk|U_k(t)|^2 \pkt
\end{equation}
The resulting equation of motion for the classical field
$\phi(t)$ is
\be
\label{eqmot}
\ddot{\phi}(t)+m^2\phi(t)+\frac{\lambda}{6}
\phi^3(t)+\frac{\lambda}{2}\phi(t){\cal F}(t)=0,
\ee
where we have introduced fluctuation integral 
\begin{equation}\label{dmsq}
{\cal F}(t)
\equiv \intk
|U_k(t)|^2 \pkt
\end{equation}
It determines the back reaction of the fluctuations onto the
classical field $\phi(t)$.

We further consider the energy density and the pressure.
The energy density is given by
\begin{equation}
\label{energie}
{\cal E}=\frac{1}{2}\dot{\phi}^2(t)+V(\phi(t))+\frac{{\rm Tr}
{\cal H}\rho(0)}{{\rm Tr} \rho(0)} \; .
\end{equation}
Calculating the trace over the Hamiltonian for the same initial state
 we obtain
\begin{eqnarray} \label{E_unren}
{\cal E}&=&\frac{1}{2}\dot{\phi}^2(t)+
\frac{1}{2}m^2\phi^2(t)+\frac{\lambda}{4!}\phi^4(t) \nonumber \\
&&+\intk\left\{\frac{1}{2}|\dot{U}_k(t)|^2
+\frac{1}{2}\omega_k^2(t)|U_k(t)|^2\right\} \pkt
\end{eqnarray}
Using the equations of motion it is easy to see that
the time derivative of the energy density vanishes.

The pressure is given by
\be
\label{pressure}
p=\dot\phi^2(t)
+\intk\left\{|\dot 
U_k^+(t)|^2+\frac{\bfk^2}{3}|U_k^+(t)|^2
\right\}-{\cal E}\; .
\ee
While the pressure does not enter the dynamics here,
so it does in FRW cosmology.

The expressions for the fluctuation integral, the
energy density and the pressure are divergent and
one has to discuss the renormalization of this
theory. 
We have presented recently \cite{Baacke:1997a}
 a fully renormalized framework
for nonequilibrium dynamics.
The main technical ingredient of this analysis is the 
perturbative expansion of the functions $U_{k}(t)$ with 
respect to orders in the potential $V(t)$.
We write the functions $U_{k}$ as
\be
U_{k}(t)=e^{-i\omk t}\left[1+h_{k}(t)\right]
\ee
and expand further in orders of the potential $V(t)$
as 
\be
h_k(t)=\sum_{n=1}^\infty h^{(n)}_{k}(t) \pkt
\ee
We also introduce the partial sums
\be
h_k^{\overline{(n)}}(t)=\sum_{l=n}^\infty h^{(l)}_{k}(t) \kma
\ee
so that
\be
h_k(t)\equiv h_k^{\overline{(1)}}(t)
=h_k^{(1)}+h_k^{\overline{(2)}}(t)\pkt
\ee
The integral equation for the function $h_k(t)$ can be derived
in a straightforward way from the 
differential equation satisfied by the functions $U_k(t)$;
it reads
\be
h_k(t)=\frac{i}{2\omk}\int_0^t dt' \left(e^{2i\omk(t-t') }-1\right)
V(t')\left[1+h_k(t')\right]\pkt
\ee
We obtain
\be
h_k^{(1)}=\frac{i}{2\omk}\int_0^t dt' 
\left(e^{2i\omk(t-t') }-1\right)V(t')\pkt
\ee
Using integrations by parts 
this function can be analyzed with respect to orders in
$\omk$ via
\bea \nonumber
h_k^{(1)}(t)&=&\frac{-i}{2\omk}\int_0^t dt' V(t')
+\sum_{l=0}^n \left(\frac{-i}{2\omk}\right)^{l+2}
\left[V^{(l)}(t)-e^{2i\omk t} V^{(l)}(0)\right]\\
&&-\left(\frac{-i}{2\omk}\right)^{n+2}
\int_0^t dt'e^{2i\omk(t-t')}
V^{(n+1)}(t') \kma \label{hexp} 
\eea
where $V^{(l)}(t)$ denotes the $l$th derivative of $V(t)$.
For energy density and pressure we need the expansion of
$\dot h_k^{(1)}(t)$ as well. From Eq. (\ref{hexp}) and the relation
\be
\dot h_k^{(1)}=2 i \omk h_k^{(1)}- \int _0^tdt'V(t')
\ee
 we find
\bea \nonumber
\dot h_k^{(1)}(t)&=&\sum_{l=0}^n \left(\frac{-i}{2\omk}\right)^{l+1}
\left[V^{(l)}(t)-e^{2i\omk t} V^{(l)}(0)\right]\\
&&-\left(\frac{-i}{2\omk}\right)^{n+1}
\int_0^t dt'e^{2i\omk(t-t')}
V^{(n+1)}(t') \pkt \label{hdotexp} 
\eea
In the following we will need the real and imaginary parts
of this expression; we introduce the following useful notation:
\bea
\calC (f,t)&=&\int_0^t dt' f(t')\cos(2\omk t') \kma\\
\calS (f,t)&=&\int_0^t dt' f(t')\sin(2\omk t') \pkt
\eea

We now insert the perturbative expansion into the fluctuation integral
to obtain
\bea  \nonumber
\calF(t)&=&\intk\left\{1+2\re h_k(t)+|h_k(t)|^2 \right\} \\
\nonumber 
&=&\intk\left\{1-\frac{V(t)}{2\omk^2}
+\frac{V(0)}{2\omk^2}\cos (2\omk t)+
\frac{\dot V(0)}{4\omk^3}\sin(2\omk t)
+\frac{\ddot V(t)}{8\omk^3}\right.\\&&\left.-\frac{\ddot V(0)}{8\omk^3}
\cos (2\omk t)-
\frac{1}{8\omk^4} \calC (\dddot V,t)
+ 2\re h_k^{\overline{(2)}}+
|h_k|^2\right\} \pkt \label{pertex}
\eea
The first two terms in the parenthesis of the second expression, i.e., 
$1$ and $V(t)/2\omk^2$ lead to divergent integrals which have
to be absorbed by the renormalization procedure. This has been
discussed in \cite{Baacke:1997a}. There, the mass $m_0$ was
chosen to be the `initial' mass $m(0)$ (see (\ref{massoft})). It was 
shown that the renormalization counter terms do not depend on this
mass but can be chosen  to contain only the renormalized mass
$m$ corresponding to the perturbative ground state at $\phi=0$. 
With this choice of initial mass
$V(0)$ is zero and the fluctuation integral is nonsingular at $t=0$. 

If, on the other hand, we choose $m_0=m$
it is obvious that the divergencies are absorbed 
 by the counter terms
depending only on the perturbative mass
$m$, but we are faced with an initial singularity arising from the
third term via (see Appendix B)
\be
\intk \frac{V(0)}{2\omk^2}\cos(2 \omk t) \simeq 
-\frac{1}{8\pi^2}\ln (2 m_0 t) \; \
{\rm as} \;\ t \to 0 \pkt
\ee
Of course, nobody has made such an `unnatural' choice 
of the initial mass, this initial singularity can be avoided trivially
by choosing $m_0=m(0)$. It is important to note, however, that
the renormalization can be performed in a way independent of the
initial condition in both cases.
The difference between the
two approaches is in the initial `vacuum' state.
These different initial states are related by a
Bogoliubov transformation
(see also Appendix A). So Bogoliubov transformations can be
used to avoid initial singularities.

In (\ref{pertex}) we have extended the expansion of
$\re h^{(1)}_k(t)$ to display also the terms of order
$\omk^{-4}$ which depend on $\dot V(0)$ and 
$\ddot V(0)$. These terms do not lead to divergencies in the 
fluctuation integral; however, in the energy and pressure
they appear multiplied with $\omk^2$ and/or $k^2$. While the energy
stays finite the pressure behaves in a singular way
via
\be
p_{\rm fluct,sing} \sim \intk 
\left[-\omk^2+\frac{k^2}{3}\right]
\left\{\frac{\dot V(0)}{4\omk^3}\sin (2\omk t)
-\frac{\ddot V(0)}{8\omk^4}\cos (2\omk t)\right\}\pkt
\ee
The behavior of the momentum integrals 
is given in Appendix B, they result in a $1/t$ singularity 
proportional to $\dot V(0)$ and a logarithmic one proportional
to $\ddot V(0)$. Therefore, these terms
have to be removed as well.


\section{Removing the initial singularity}
\setcounter{equation}{0}
We have seen in the previous section that nonzero initial
values of $V(t)$ and its derivatives lead to initial singularities.
The clue for dealing with these terms has already been indicated:
the leading singularity can be removed by a Bogoliubov
transformation from the
perturbative vacuum to a vacuum corresponding to
free quanta of the initial mass $m(0)$.
We expect, therefore, that the other singular terms can
be removed in this way as well.

We define a general initial state by requiring that
\be
\left[a(\bfk)-\rho_ka^\dagger(\bfk)\right]|i\rangle =0 \pkt
\ee
The Bogoliubov transformation to this state
is given in Appendix A.
If the fluctuation integral, the energy and the pressure are
computed by taking the trace with respect to this
state the functions $U_k(t)$ are just replaced by
\be
F_k(t)=\cosh(\gamma_k)U_k(t)+e^{i\delta_k}
\sinh(\gamma_k)U_k^*(t) \kma
\ee
where $\gamma_k$ and $\delta_k$ are defined by
the relation
\be
\rho_k=e^{i\delta}\tanh(\gamma_k)\pkt
\ee
The fluctuation integral now becomes
\bea
\calF(t)&=&\intk |F_k(t)|^2 \nonumber
\\ &=&
\intk \left\{\cosh (2 \gamma_k(t)) |U_k(t)|^2+
\sinh (2\gamma_k)\re\left( e^{-i\delta_k}U^2_k(t)\right)\right\}\pkt
\eea
Expanding as before we find
\bea \nonumber
\calF (t)&=&\intk\left\{
\cosh(2\gamma_k)
\left[1-\frac{V(t)}{2\omk^2}+\frac{V(0)}{2\omk^2}\cos(2\omk t)
+\frac{\dot V(0)}{4\omk^3}\sin(2\omk t)\right.\right.
\\ \nonumber&&\left.+\frac{\ddot V(t)}{8\omk^4}
-\frac{\ddot V(0)}{8\omk^4}\cos(2\omk t)-
\frac{1}{8\omk^4}\calC(\dddot V,t)
+ 2\re h_k^{\overline{(2)}}+|h_k|^2 \right]
\\ &&+\sinh (2\gamma_k)\cos(\delta_k)\cos(2\omk t)
\label{fluc2}
\\ \nonumber
&&-\sinh(2\gamma_k)\sin(\delta_k)\sin(2\omk t)
+\sinh(2\gamma_k)\re e^{-2i\omk t-i\delta}
\left(2h_k+h_k^2\right)  \Bigg\}\pkt
\eea
Let us first discuss how to get rid
of the most singular term, proportional to 
$V(0)$. Requiring this term to be compensated by the terms
proportional to $\sinh(2\gamma_k)$ we find
\bea 
\delta_k&=&0\kma\\
\tanh(2\gamma_k)&=&-\frac{V(0)}{2\omk^2}\pkt
\eea
As explained in Appendix A the standard Bogoliubov 
transformation from the perturbative vacuum
with mass $m$ to the vacuum corresponding to 
quanta with the initial mass $m(0)$ is 
mediated by a function $\gamma_k'(k)$ satisfying
\be
e^{\gamma'_k}=\left(\frac{m^2+\bfk^2}{m^2(0)+\bfk^2}\right)^{1/4}\kma
\ee
which implies
\be
\tanh(2\gamma'_k)=\frac{-V(0)}{2\omk^2+V(0)}
\pkt\ee
We see that $\gamma_k$ and $\gamma'_k$ agree asymptotically
to leading order in $1/\omk$. So requiring that the most 
pronounced initial singularity vanishes leads essentially to the
usual choice for the initial
state, namely $m_0=m(0)$ and therefore $V(0)=0$ .
The analysis of subleading terms in the difference between
$\gamma_k$ and $\gamma'_k$ becomes somewhat cumbersome.
After we have convinced ourselves that the Bogoliubov transform
is the right technique for getting rid of initial singularities we will
therefore choose $m_0=m(0)$ as everybody does and apply this technique
to get rid of the remaining singularities. So from now on $V(0)=0$
and $\omk=(\bfk^2+m^2(0))^{1/2}$.
Requiring that the terms proportional to
$\dot V(0)$ and $\ddot V(0)$ vanish leads to the 
conditions 
\bea \label{deltadef}
\tan(\delta_k)&=&2\omk\frac{\dot V(0)}{\ddot V(0)}\kma
\\ \label{gammadef}
\tanh (2\gamma_k)&=&
\frac{1}{4\omk^3}\left[\dot V^2(0) +\frac{\ddot V^2(0)}
{4\omk^2}\right]^{1/2} \pkt
\eea
Using these functions we are now ready to formulate the
renormalized equation of motion and the energy momentum tensor.


\section{The renormalized equations}

We have given the bare equation of motion and energy momentum
tensor in section 2. The renormalization for the original
initial state has been discussed in \cite{Baacke:1997a}. 
We have to ensure now that the scheme used there is not spoiled
by the improved initial state. The main new feature
in the fluctuation integral, the energy density and the
pressure is the appearance of factors $\cosh(2\gamma_k)$ and
$\sinh(2\gamma_k)$. We will need their asymptotic behavior.
Using Eq. (\ref{gammadef}) we have
\be
\gamma_k\stackrel{k\to\infty}{\simeq}\frac{|\dot V(0)|}{8\omk^3}
\pkt\ee
The factor $\cosh(2\gamma_k)$ is equal to $1$ for $\gamma_k=0$;
we will need the difference
\be \label{cosest}
\cosh(2\gamma_k)-1=2\sinh^2 (\gamma_k)\stackrel{k\to\infty}{\simeq}
\frac{\dot V^2(0)}{32\omk^4}\pkt
\ee
There are new terms proportional to $\sinh(2\gamma_k)$; this
factor behaves as
\be \label{sinest}
\sinh(2\gamma_k)\stackrel{k\to\infty}{\simeq}
 \frac{|\dot V(0)|}{4\omk^3}\pkt
\ee

The dimensionally regularized fluctuation integral (\ref{fluc2})
takes, after cancellation of the singular integrals induced
by Eqs. (\ref{deltadef}) and (\ref{gammadef}),
the form
\bea \nonumber
\calF_{\rm reg} (t)
&=&-\frac{m_0^2}{16\pi^2}(L_0+1)
-\frac{V(t)}{16\pi^2}L_0
\\ &&+\intk\left\{
\sinh^2(\gamma_k)\left[1-\frac{V(t)}{2\omk^2}\right]\right.
\\&&+\cosh(2\gamma_k)\left[\frac{\ddot V(t)}{8\omk^4}
\frac{1}{8\omk^4}\calC(\dddot V,t)
+ 2\re h_k^{\overline{(2)}}+|h_k|^2 \right]
\\ \nonumber
&&+\sinh(2\gamma_k)\re e^{-2i\omk t-i\delta}
\left(2h_k+h_k^2\right)  \Bigg\}\pkt
\eea
Here we have introduced the abbreviation
\be
L_0=\frac{2}{\epsilon}+\ln \frac{4\pi \mu^2}{m_0^2}-\gamma \pkt
\ee
Using the estimates (\ref{cosest}) and (\ref{sinest}), and using the
fact that the mode function $h_k$ behaves as $\omk^{-1}$ we see that
the momentum integral is convergent.

Introducing the counter term Lagrangian
\be
{\cal L}_{\rm c.t.}= \frac{1}{2}\delta m^2 \Phi^2
+\frac{\delta\lambda}{4!} \Phi^4
\ee
the fluctuation integral gets replaced, in the equation of motion
(\ref{eqmot}) for $\phi(t)$, by
\be
\calF_{\rm fin}=\calF_{\rm reg}+ \frac{2\delta m^2}{\lambda} 
+ \frac{\delta \lambda}
{3\lambda}\phi^2(t)\pkt
\ee
With the standard choice
\bea
\delta m^2 &=&
\frac{\lambda m^2}{32\pi^2}(L+1)\kma
\\
\delta \lambda &=&\frac{3\lambda^2}{32\pi^2}L\kma
\\
L&=&\frac{2}{\epsilon}+\ln \frac{4\pi \mu^2}{m^2}-\gamma
\eea
$\calF_{\rm fin}$ is indeed finite.

The calculation of the energy density proceeds in an analogous way.
The fluctuation energy becomes, using dimensional regularization,
\bea
\calE_{\rm fluct} &=&
\frac{1}{2}\intk\left\{|\dot F_k(t)|^2+(\omk^2+V(t))|F_k(t)|^2\right\}
\\ \nonumber
&=&-\frac{m_0^2}{64\pi^2}(L_0+\frac{3}{2})
-\frac{V(t)}{32\pi^2}(L_0+1)-\frac{V^2(t)}{64\pi^2}L_0
\\ \nonumber 
&&+\frac{1}{2}\intk \Bigg \{ 2 \sinh^2(\gamma_k)
\left[2\omk^2+V(t)\left(1-\frac{V(t)}{2\omk^2}\right)\right]\\
\nonumber
&&+V(t)\cosh(2\gamma_k)\left[\frac{\ddot V(t)}{8\omk^4}
-\frac{\calC(\dddot{V},t)}{8\omk^4}+
2 \re h_k^{\overline{(2)}}+|h_k|^2\right]
\\ \nonumber
&&+V(t)\sinh(2\gamma_k)\re e^{-2i\omk t-i\delta_k}\left(2h_k
+h_k^2\right)
\\ 
&&+\cosh(2\gamma_k) |\dot h_k|^2
\\ \nonumber
&&+\sinh(2\gamma_k)\re e^{-2i\omk t -i\delta_k}
\left[ \dot h_k^2-2i\omk (1+h_k)\dot h_k\right]\Bigg\}\pkt
\eea
The divergent parts are cancelled by the counter terms
\be
\calE_{c.t.}=\delta \Lambda+\frac{1}{2}\delta m^2 \phi^2(t)
+\frac{\delta\lambda}{4!}\phi^4(t)
\ee
with the `cosmological constant' counter term
\be
\delta \Lambda = \frac{m^4}{64\pi^2}(L+\frac{3}{2})\pkt
\ee

Finally, we have to consider the pressure. We find
for the regularized fluctuation part:
\bea\nonumber
p_{\rm fluct}&=&-\calE_{\rm fluct}
-\frac{m_0^4}{96\pi^2}-\frac{V(t)}{48\pi^2}m_0^2-
\frac{\ddot V(t)}{96\pi^2}(L_0+\frac{1}{3})
\\ \nonumber&&
+\intk\Bigg\{
\sinh^2(\gamma_k)\left[
2\omk^2+\big(-\omk^2+\frac{k^2}{3}\big)
\left(1-\frac{V(t)}{2\omk^2}+\frac{\ddot V(t)}{8\omk^4}\right)\right]
\\ 
&&+\cosh(2\gamma_k)\left[-\frac{\calC(\dddot{V},t)}{8\omk^4}
+2 \re h_k^{\overline{(2)}}+|h_k|^2\right] 
+\cosh(2\gamma_k)|\dot h_k|^2 
\\ \nonumber
&&\sinh(2\gamma_k)\re e^{-2i\omk t-i\delta_k} 
\left[\big(-\omk^2+\frac{k^2}{3}\big)
\left(2 h_k +h_k^2\right)+\dot h_k^2
-2i\omk(1+h_k)\dot h_k\right]
\Bigg\}\pkt
\eea
In order to cancel the divergent term 
proportional to $\ddot V$ one introduces
a counter term for the energy momentum tensor
\be
\delta T_{\mu\nu}=A(g_{\mu\nu}\partial_\alpha\partial^\alpha
-\partial_\mu\partial^\nu)\Phi^2\kma
\ee
which leads to a counter term
\be
p_{c.t.}=A \frac{d^2}{dt^2}\phi^2(t)=\frac{2A}{\lambda}\ddot V(t)
\pkt\ee
in the pressure.
We choose
\be
A = -\frac{\lambda}{192\pi^2}L
\pkt\ee
The remaining momentum integral is finite and nonsingular
in $t$. For most of the terms this can
be seen by inspection using Eqs. (\ref{cosest}), (\ref{sinest}), and
the expansions (\ref{hexp}) and (\ref{hdotexp}). There are some internal
cancellations which are, however, the same as for the case
$\gamma_k=0$ already discussed in \cite{Baacke:1997a}.
The only new, potentially singular term is
\be
\intk\sinh(2\gamma_k)\big(-\omk^2+\frac{k^2}{3}\big)
\re e^{-2i\omk t}2 h_k\pkt
\ee
The leading singular behaviour is given by
\be
\dot V(0)\int_0^tdt'V(t')
\intk\frac{1}{4\omk^4}\big(-\omk^2+\frac{k^2}{3}\big) \cos (2\omk t)
\pkt\ee
While the momentum integral behaves as $\ln t$ as $t\to 0$ the
time integral behaves as $t^2$ since $V(0)=0$.
So the renormalized pressure is indeed nonsingular at $t=0$.

From the analysis of divergent integrals given in this section it
is obvious that only the leading asymptotic behavior of
$\gamma_k$ is relevant, more precisely, only the terms of order 
$\omk^{-3}$ and $\omk^{-4}$. This means that any Bogoliubov
transformation whose function $\gamma_k$ has this leading asymptotic
behavior is equally suitable for defining an appropriate
initial state.

We have formulated our modified renormalized equations 
for $\lambda \Phi^4$ theory in flat space. The generalization
to a scalar field in a flat
Friedmann-Robertson-Walker universe
is straightforward and the cancellation of singular
terms in the energy density and the trace of the energy momentum
tensor proceeds in the same way. 

\section{Conclusions}

We have considered here the choice of initial states for
a nonequilibrium system in quantum field theory.
Our considerations arose from the problem that
logarithmic and linear singularities in the variable
$t-t_0$ appear in the energy momentum tensor 
and affect the dynamics
of FRW cosmology. We consider such singularities
- and their consequences - as unphysical,
at least if $t_0$ is just some conveniently chosen point in time  
within a continuous evolution of the system.   
Most authors, including ourselves, have chosen initial
states that correspond to equilibrium states of the system.
We have constructed here improved initial states
for nonequilibrium systems in such a way that
the appearance of initial singularities is avoided.
These states are obtained from the usual `vacuum' states
by a Bogoliubov transformation. The essential part of
this transformation is a Bogoliubov `rotation' of the
creation and annihilation operators at large momentum.
The construction presented here specifies a transformation
of creation and annihilation operators at all momenta.
It is not unique in the sense that it may be arbitrarily
modified at small momenta. This non uniqueness is, however,
nothing else as the freedom for choosing an 
`arbitrary', pure or mixed,  initial state. 
Our construction
can be considered as formulating a minimal requirement 
for choosing such states in the sense that it 
specifies the initial
state of the high momentum quantum modes.


\section*{Acknowledgements}
It is a pleasure to thank H. de Vega and A. di Giacomo
for discussions on the problem.
\begin{appendix}


\section{ Bogoliubov transformation}
\setcounter{equation}{0}
In this Appendix we briefly recall some basic
features of the Bogoliubov transformation (see , e.g., 
\cite{Bogoliubov:1983}).
We start with a vacuum state defined by
\be
a(\bfk)|0\rangle =0 \pkt
\ee
We would like to obtain a new state $|\tilde 0\rangle$ that is
annihilated by $a(\bfk)+\rho_ka^\dagger(\bfk)$, where $\rho_k$ is some
complex function of $k$, i. e. we require
\be \label{newvac}
\left[a(\bfk)-\rho_ka^\dagger(\bfk)\right]|\tilde 0\rangle =0 \pkt
\ee
Such a state can be obtained from $|0\rangle$ by a 
Bogoliubov transformation
\be
|\tilde 0\rangle = \exp(Q)| 0\rangle \pkt
\ee
Using the general relations given in \cite{Bogoliubov:1983} one finds
the explicit form of the operator $Q$ as
\be
Q =\frac 1 2 \intk \gamma_k\left[e^{i\delta_k}
a^\dagger(\bfk)a^\dagger(-\bfk)
-e^{-i\delta_k}a(\bfk)a(-\bfk)\right] \pkt
\ee 
Here $\gamma_k$ and $ \delta_k$ are defined by the relation
\be
\rho_k=e^{i\delta_k}\tanh \gamma_k
\ee
so that (\ref{newvac}) can also be written as
\be
\tilde a(\bfk) |\tilde0\rangle
=\left [\cosh( \gamma_k) a(\bfk) + e^{i\delta_k} \sinh (\gamma_k)
a^\dagger(-\bfk)
\right] |\tilde 0\rangle =0  \pkt
\ee
A special class of new `vacuum' states $|\tilde0\rangle$ is obtained
when the new creation and annihilation operators refer to free particles
with a different mass $\tilde m_0$. In the field
expansion (\ref{fieldexp}) this
means that the energy $\omk=(\bfk^2+m_0^2)^{1/2}$ is replaced
by $\tilde \omk=(\bfk^2+\tilde m_0^2)^{1/2}$. In this case
\be
\tilde a (\bfk) =\sqrt{\frac{\tilde\omk}{\omk}}
a(\bfk)+\sqrt{\frac{\omk}{\tilde\omk}} a^\dagger(\bfk)
\ee
and therefore
\be
\gamma_k=\frac 1 2 \ln \frac{\omk}{\tilde\omk}    
\ee
while $\delta_k=0$.
For $k \gg m_0,\tilde m_0$ the function $\gamma_k$ behaves as
\be
\gamma_k \simeq \frac{m_0^2-\tilde m_0^2}{4 k^2}\pkt
\ee


\section{Some singular integrals}
\setcounter{equation}{0}
The singular behaviour in time arises from the following integrals
\be
I_1(t)=\intk \frac{1}{\omk^2} \cos(\omk t)
\ee
and 
\be
I_2(t)=\intk \frac{1}{\omk} \sin(\omk t)\pkt
\ee 
The first integral can be rewritten as
\bea \nonumber
I_1(t)&=&\frac{1}{4\pi^2}
\int_m^\infty\frac{d\omega\sqrt{\omega^2-m^2}}
{\omega^2}\cos(2\omega t)\\&=&
\frac{1}{4\pi^2}
\int _m^\infty d\omega \left(\frac{1}{\sqrt{\omega^2-m^2}}
\cos(2\omega t) - 
\frac{m^2}{\omega^2\sqrt{\omega^2-m^2}}
\cos(2\omega t)\right)\pkt
\eea
The integral over the second term is nonsingular; the first term
yields a Bessel function $Y_0(2 m t)$, explicitly
\be
I_1(t) = -\frac{1}{8\pi}Y_0(2 m t) + O(t^2) 
\stackrel{t\to 0}{\simeq}
-\frac{1}{4\pi^2}\ln(2mt)\pkt
\ee
The integral $I_2(t)$ is simply given by
\be
I_2(t)=-\frac{1}{2}\frac{d}{dt}I_1(t)
\ee
and therefore
\be
I_2(t)\stackrel{t\to 0}{\simeq} \frac{1}{8\pi^2 t}\pkt
\ee
\end{appendix}

\end{document}